# What model(s) for program understanding?

**Françoise Détienne**
Projet de Psychologie Ergonomique
INRIA
Domaine de Voluceau, Rocquencourt,
BP 105, Rocquencourt
78153 Le Chesnay cedex
France
Phone: 33 1 39 63 55 22
Fax: 33 1 39 63 53 30
Email: Francoise.Detienne@inria.fr

**Abstract**

The first objective of this paper is to present and discuss various types of models of program understanding. They are discussed in relation to models of text understanding. The second objective of this paper is to assess the effect of purpose for reading, or more specifically programming task, on the cognitive processes involved and representations constructed in program understanding. This is done in the theoretical framework of van Dijk and Kintsch's model of text understanding (1983).

# 1. Introduction

A computer program is a procedure which solves a problem and which is coded in a highly constrained language, a programming language. Program representation is a very complex multidimensional representation. In this way it is close to procedure representation as formalised by Baudet and Cordier (1995). A program may be represented as a hierarchical tree of goals/subgoals, with objects represented as central or secondary entities and with plans which achieve the goals. Relationships such as data flow, control flow, and plan structures may also be represented.

The first objective of this paper is to present and discuss various types of models which seek to account for program understanding. They are discussed in relation to models of text understanding. The second objective of this paper is to assess the effect of purpose for reading, or more specifically programming task (e.g. modifying a program), on the cognitive processes involved and representations constructed in program understanding. This is done in the theoretical framework of van Dijk and Kintsch's model of text understanding (1983). In this approach, which we refer to as the mental model approach, the distinction is made between two kinds of mental representations, the textbase and the situation (or mental) model. Our position is that purpose for reading has an impact on the cognitive processes involved, and on the kind of representation constructed, i.e., a textbase representation versus a mental model. This position is defended on the basis of results of empirical studies on software documentation, modification and reuse.

# 2. Models of text understanding

Various models have been developed to account for the understanding of texts written in natural language. We will roughly distinguish the functional approach, the structural approach and the mental model approach. These approaches differ partly according to the direction of the processes, top-down versus bottom-up, and the use of content versus structural knowledge.

According to the <u>functional approach</u> (Galambos, Abelson & Black, 1986; Schank & Abelson, 1977), the orientation of the processes is mostly top-down and the processes are content-oriented. The reader evokes content schemas, e.g., scripts, goals, plans. These activations allow the reader to make inferences and expectations.

Two types of models belong to the <u>structural approach</u>: the structured-schema approach and the propositional-network approach. The structured-schema approach (Meyer, 1975; Mandler & Johnson, 1977) shares a top-down orientation with the functional approach. However, this model is strictly structural. Structural schemas are activated and guide the understanding process. The representation constructed is formed by propositions which are connected by structural links. The propositional-network approach (Norman & Rumelhart, 1975; Kintsch, 1974) is also structural but, in contrast to the former approach, the orientation of the understanding processes is mostly bottom-up. The representation constructed is formed by propositions which are connected by referential links.

The <u>mental model approach</u> corresponds to a recent evolution of van Dijk and Kintsch's model (Kintsch, 1988; van Dijk & Kintsch, 1983) which takes into account the role of content knowledge in text understanding. The model is no longer strictly structural but rather combines the structural and functional approaches. Three distinct, but interacting, levels of cognitive representation, are distinguished:

level 1. the surface form representation,
level 2. the propositional textbase representation
level 3. the situation model or mental model

Levels 1 and 2 correspond to linguistic representations of the text which are isomorphic to the text structure. They reflect what is contained in the text at a surface level and at a propositional level. Level 3 corresponds to an a-linguistic representation of the text and reflects the world situation referred to by the text. It is initially built up from a linguistic representation and makes extensive use of the subject's existing domain knowledge. It is produced by inferences and is also a source for making new inferences.

## 3. Models of program understanding

Program understanding has been extensively studied and there is some debate between the advocates of two classes of models. Some authors argue that text understanding models may account for program understanding whereas other authors argue that problem solving models are more relevant.

### 3.1 Text understanding approaches

The three text understanding approaches presented above have been followed to account for program understanding, i.e., the functional approach, the structural approach, and the mental model approach.

According to the functional approach, program understanding is assumed to correspond to processes of schema activation and instantiation (Black, Kay, & Soloway, 1986; Détienne, 1988; 1990; Détienne & Soloway, 1990; Soloway, Ehrlich, & Bonar, 1982; Wiedenbeck, 1986). Schemas are knowledge structures stored in memory which represent generic structures like sequences of steps for solving a problem. Many studies have focused on identifying and formalising schemas possessed by expert programmers. Studies have lent empirical support to this approach.

According to the structural approach, program understanding corresponds to the construction of a propositional network. This construction is made by bottom-up processes. A few authors (Atwood & Ramsey, 1978, Vessey, 1989) have followed this approach. The propositional hierarchy is defined relatively to the embedding levels of the control structure. In a debugging task, these authors assumed that the deeper a bug was in the propositional hierarchy, the harder it would be to locate it[1]. The data provided little empirical evidence to support this approach.

According to the mental model approach, program understanding corresponds to constructing a representation of the situation. This approach has been followed to account for procedural programs understanding (Pennington, 1987a, 1987b). Pennington distinguishes between the program model (or textbase) which reflects the text-based representation of the program (i.e. elementary operations and control flow) and the domain model (or mental model) which reflects the entities of the problem domain and their relationships (i.e. functions/goals of the problem and data flow). This model has a bottom-up orientation. The textbase representation is assumed to emerge first during program comprehension.

---

[1] It was also assumed that constructing the propositional network would precede the debugging activity based on diagnosis reasoning.

This approach to program understanding has received some empirical support. Pennington (1987a) found that after reading short programs, expert programmers answered questions on control flow more correctly than questions on functions. The author interprets this result as supporting the hypothesis that the textbase representation emerges first during program comprehension. Then the programmers had to perform a modification task on the program they had read. After having performed the task, they answered questions on functions (or problem goals) more correctly. The author interprets this result as supporting the hypothesis that, with a modification task orientation, programmers construct a mental model.

More recently, the mental model approach has been followed to account for object-oriented program understanding (Détienne, Burkhardt & Wiedenbeck, 1996). The textbase representation is assumed to be composed of: (1) elementary operations at a micro-level, (2) elementary functions (routines) attached to objects and control flow information at a Macro level. The textbase reflects the program structure, i.e. a program is structured in routines performing functions, and a type of link explicit in the program text, i.e. the control flow. The mental model is assumed to be composed of three viewpoints, i.e., the object viewpoint, the functional viewpoint, and the communicationnal viewpoint. The construction of the mental model is made by inferences based on information from the textbase and on knowledge activated in memory: schemas and episodic knowledge in the problem domain as well as in the programming domain. In contrast to Pennington, these authors do not adopt a strictly bottom-up approach.

### 3.2 Problem solving approach

According to the problem solving models, program understanding corresponds to <u>problem solving and plan recognition mechanisms</u> (Koenemann & Robertson, 1991; Robertson, Davis, Okabe & Fitz-Randolf, 1990). These authors highlight the importance of selection processes and selective representation in program understanding. An argument in favour of this approach is that programmers read the code in a non linear order which would reflect decision making and, more generally, reasoning processes involved in reading. Robertson et al. (1990) analysed the reading strategies of experienced programmers. They found that 11% of the reading activity involved switches in direction. 17% of the search activity involved going backwards through the code. Switching times were relatively longer which was interpreted by the authors as reflecting decision making.

Authors defending the problem solving approach oppose this approach to the purely functional and purely structural approaches to program understanding. They do not discuss the mental model approach. However the mental model approach accounts for mechanisms proposed by the problem solving approach in terms of inferences produced while constructing the mental model and selective encoding processes. Furthermore authors defending the problem solving approach have a simplistic view of understanding models because they assume that these models do not take into account strategic variations related to the task. This is in marked contrast to the mental model approach as discussed below.

## 4. Effect of Purpose for reading

An important issue is the effect that purpose for reading has on program understanding. A program is usually read with a particular goal in mind: modification, reuse, debugging or documenting. Our point is that assessing what type of model accounts for program understanding can only be done by taking into account the effect of purpose for reading. The mental model approach takes into consideration the effect of task on text understanding. This is why we have chosen this theoretical framework in order to discuss the effect of programming task on program understanding. We first review some results on the effect of purpose for reading on text understanding in the mental model approach. Then, taking this theoretical framework as a basis, we discuss results on program understanding oriented by various tasks.

### 4.1 Effect of purpose for reading on text understanding

The mental model approach, in its recent evolution, takes into account the effect of purpose for reading on text understanding. Two main categories of purposes for reading (Mannes, 1988; Mills, Diehl, Birkmire & Mou, 1995; Richard, 1990; Schmalhofer & Glavanov, 1986) are distinguished : read-to-recall versus read-to-do. Studies have provided evidence that these two types of goals have distinct effects on the encoding processes and on the type of representation which is constructed in text understanding.

One hypothesis is that the <u>read-to-recall purpose for reading</u> focuses the understanding activity on the construction of the textbase (what is said, and how it is said) whereas the <u>read-to-do purpose for reading</u> focuses the understanding activity on the construction of the mental model (what the situation of reference is). Mills et al. (1995) showed that read-to-recall participants recall a procedural text better whereas read-to-do participants perform the task (described by the text) better. Read-to-do participants recall less information judged to be less important for performing the task than do read-to-recall participants. These results provide empirical support to the hypothesis above. Text summarisation and knowledge acquisition tasks have been used by Schmalhofer and Glavanov (1986) to study the effect of read-to-recall and read-to-do purposes. In this study, subjects had to read a programmer's manual for one of these two purposes for reading. The authors found that the subjects who studied for text summarisation remembered more propositional information while subjects with a knowledge acquisition goal remembered more situational information.

Another hypothesis is that differential encoding processes are involved depending on purpose for reading. In Mills et al.'s study (1995) it was found that reading rates varied as a function of high and low importance of the information regarding task performance and varied more for the read-to-do participants than for the read-to recall participants. In Schmalhofer and Glavanov's study (1986) it was found that the reading time patterns were different according to the goal of subjects. The knowledge acquisition subjects read faster and had a different pattern of reading times than the text summarisation subjects, as would be expected since summarisation subjects had to concentrate on all the information while the knowledge acquisition subjects did not. These results confirm the hypothesis on differential encoding processes.

## 4.2 Effect of purpose for reading on program understanding

In the mental model approach, purpose for reading has an effect on the encoding process and on the type of representation constructed, i.e., textbase versus mental model. We adopt this approach to discuss results on program understanding oriented by various tasks. In this theoretical framework, we will distinguish the effect of purposes for reading discussed above, i.e., read-to-recall and read-to-do.

4.2.1 <u>Effect of Read-to-recall purpose</u>

Reading a program for documenting it can be considered similar to the read-to-recall task (e.g., text summarisation) for text understanding. Similarly to the text summarisation task, programmers who read a program for documenting it should concentrate on encoding the program text itself, i.e., constructing a textbase representation. The task of documenting a program produced by someone else has been studied by Rouet, Deleuze-Dordron and Bisseret (1994). A general hypothesis was that comments reflect the designer's cognitive representation of the entity being commented. The analysis of comments elicited several categories of information: paraphrases (comments paraphrase program statements and do not include any new information.), syntactic explanations (about programming rules), semantic explanations (about solutions being implemented), meta-comments (statements about commenting) and inferences from labels. Experts issued mostly explanations, then paraphrases. This result suggests that the representation constructed in a documentation task reflects (1) low level functional information close to Pennington's elementary operation category (called semantic explanation by the authors), and (2) control flow information (paraphrases). This suggests that programmers have constructed a textbase[2] representation. Another result was that "structural" units ,e.g., beginnings of loops, were the most frequently commented. This suggests that the structure of the representation constructed reflects the structure of the program text (as defined by the control structure).

Another study on documentation (Riecken, Koenemann-Belliveau & Robertson, 1991) showed that expert programmers documenting a program produced by someone else generated more comments detailing given instructions explicitly stated in the code rather than general domain information associated with the task. Subjects generated nearly twice as many detailed comments as abstract comments. These results suggest again that programmers in a documentation task construct a textbase representation rather than a mental model. Furthermore, it was found that subjects located vertical spacing according to the program text structure, e.g., between routines. These last results also support the textbase construction hypothesis since the constructed representation preserves the text structure.

Another example of the effect of read-to-recall purpose for reading is presented by Pennington (1987a). In her first study, the programmers were instructed that they had to read a program in order to answer questions that would be asked later on. In the first phase of her second study, they were instructed to read a program in order to make a modification task later on. However the modification task was specified only in a second phase. After these first phases subjects were asked different kinds of question about the program. In both studies, it was found that expert programmers answered questions on control flow more correctly than questions on functions/goals. These results again support the textbase construction hypothesis. It should be remarked that subjects in the "still unspecified" modification task (study 2) constructed the same kind of representation as in a purely read-to-recall task (study 1).

---

[2] Constructing a textbase is likely to be implied in program documentation as far as we consider the task of documenting a program produced by someone else. However, Rouet, Deleuze-Dordron and Bisseret (1995) showed that documenting may be part of the design activity itself when documenting is involved during the design of one's own program. In this case the representation constructed is no longer restricted to a textbase.

4.2.2 <u>Effect of Read-to-do purpose</u>

The effect of the read-to-do purpose for reading may be examined in various tasks, e.g., program modification, and program reuse. In these situations the programmers read a program in order to use it for performing a task. In the modification situation, the task is to take into account new specifications. The read program has to be modified in order to meet new constraints or goals of the problem. In the reuse situation, the task is to design a program. The read program is reused in order to design or implement another software program. Two main results emerge from studies program modification and program reuse. As in text understanding studies, the read-to-do task has an effect on the encoding processes and entails the construction of a mental model.

Several studies (Littman, Pinto, Letovsky & Soloway, 1986; Koeneman & Robertson, 1991) on the modification task show that differential encoding processes are involved. It is found that programmers use as-needed strategies. They study code or documentation only if they believe that the code is relevant for the task. Koeneman and Robertson (1991) distinguish between three levels of relevance: direct relevance, i.e., code segments that have to be modified, intermediate relevance, i.e., code segments that are perceived to interact with relevant code, and strategic relevance, i.e., code which serves to locate or detect directly or intermediate relevant code. As noted by the authors (p 129): "the [modification] task on hand determines the scope and focus of attention. For one modification it might be sufficient to know how a piece of code works while for a different modification the question of why this implementation was chosen is of great importance."

Read-to-do purpose for reading focuses the understanding activity on the construction of the mental model. Studies on software modification and on software reuse provide empirical support to this hypothesis. In Pennington's study (1987a), programmers had to perform a modification task after having read the program in a first phase. After having performed the modification, they answered questions on functions (or problem goals) more correctly than questions on control flow. The author interprets this result as supporting the hypothesis that, with a modification task orientation, programmers construct a mental model. However, the experimental procedure did not allow a distinction to be made between the effect of task orientation (or purpose for reading) and the effect of extra time spent reading the program.

Several studies on software reuse (Burkhardt & Détienne, 1995; Rouet, Deleuze-Dordron & Bisseret, 1995) show that, when a source component is evoked or retrieved in a problem solving phase (as opposed to an implementation phase) of software design, information about the source situation from which the component comes from is searched for or inferred. Programmers infer solution goal structure, constraints, evaluation criteria or design rationales. In this case, it seems that reusing a component implies more than constructing a textbase representation of the source component itself. It implies constructing a mental model of the source situation. This mental model allows the representation constructed for solving the design problem on hand to be enriched and the search space to be enlarged.

## 5. Conclusion

To sum up, the mental model approach is an interesting theoretical framework to study program understanding. One of the main interests is that it takes into account the effect of the task on the understanding activity. Our last point will be to briefly discuss the notion of purpose for reading. In a first step we have matched it to the notion of programming task, e.g., software modification. Similarly to the opposition made between the notions of prescribed task and effective task (Leplat & Hoc, 1983), it may be important to distinguish between the notions of prescribed purpose for reading and effective purpose for reading. This latter notion would refer to the representation of purpose for reading constructed by a programmer or, more generally, a reader. This construction is probably influenced by various factors, internal and external to the programmer (e.g., subject's expertise, programming environment), as well as possible interactions between these factors.

## Acknowledgements

Special thanks to Willemien Visser for her comments on a previous draft of this paper.